\documentclass[aps,twocolumn]{revtex4}
\usepackage{amsfonts,amssymb}
\input{epsf}
\def\beq{\begin{eqnarray}}
\def\eeq{\end{eqnarray}}
\def\nn{\nonumber\\}

\def\ben{\begin{enumerate}}
\def\een{\end{enumerate}}
\def\bem{\begin{itemize}}
\def\eem{\end{itemize}}

\def\half{{\textstyle{1\over2}}}

\begin{document}
\title{Counting black hole microscopic states in loop quantum gravity}


\author{A. Ghosh}\thanks{amit.ghosh@saha.ac.in}
\author{P. Mitra}\thanks{parthasarathi.mitra@saha.ac.in}
\affiliation{Theory Division, Saha Institute of Nuclear Physics\\
1/AF Bidhannagar, Calcutta 700064}
\begin{abstract}
Counting of microscopic states of black holes is performed within the framework of
loop quantum gravity. This is the first calculation of the pure {\em horizon} states
using statistical methods, which reveals the possibility of additional states missed
in the earlier calculations, leading to an increase of entropy. Also for the first
time a microcanonical temperature is introduced within the framework.
\end{abstract}
\pacs{04.70.Dy, 04.60.Pp}
\maketitle
\vskip .5in


\section{Introduction}
A nonperturbative framework of quantum gravity using holonomy as fundamental
variables, popularly known as loop quantum gravity, has been in vogue for some years
now, see \cite{ash0} for a recent survey. In this framework, a start was made in
\cite{ash} in the direction of quantizing a black hole and thereby counting its
microstates. In this approach, a black hole is characterized {\em effectively} by an
isolated horizon, see \cite{ashiso} and references therein. The quantum states arise
from quantizing the phase space of an isolated horizon whose cross sections, which
are two-spheres, are punctured by suitable spin networks. The spin quantum numbers
$j,m$, which characterize the punctures, also label the quantum states. The entropy
is obtained by counting the manifold possibilities of such quantum states, or
essentially the labels, that are consistent with a fixed area of the cross section
\cite{ash}.

A calculation of microscopic states was carried out in \cite{meissner} using
a recursion relation technique. Soon after, in \cite{gm}, an explicit combinatorial
method was introduced, which in addition to counting states also gives the dominant
configuration of spins, namely the configuration yielding the maximum number of
states. However, the two counting calculations gave slightly different results. See
also \cite{cor} for a recent survey, which supports the result of \cite{gm}. The
root of this difference has been briefly discussed in \cite{gm}:
while \cite{meissner} takes into account only the spin projection ($m$) labels of
the microstates, thus counting what we refer to as the pure {\em horizon} states 
in the present work, \cite{gm} and \cite{cor} 
take into
account the spin $j$, which in some sense characterizes the bulk states, as well as
the $m$-labels. Any counting involves two constraints to be met. While one of them,
the spin projection constraint (see below for details), which arises from an
interplay of the bulk and the horizon Hilbert spaces, can be expressed solely in
terms of the $m$-labels, the other constraint involving the area of the horizon,
cannot bypass the $j$-labels. The calculation of \cite{gm} was essentially based
upon the intuition that a quantum isolated horizon can never be completely
characterized by states of the horizon (or surface) Hilbert space, the bulk states
play an essential r\^{o}le. 

The first part of the present work uses the {\em combinatorial
method} of \cite{gm} to count the number of the pure {\em horizon} states
which were sought to be counted in \cite{meissner}. This leads to an increased number. 
Unlike the result of \cite{meissner}, this number is
consistent with the thesis presented in \cite{dreyer}, in the sense that $j=1$
allows three values of $m$.
In the second part of the work, we introduce a microcanonical temperature for 
each null normal vector field defined on the horizon. It involves the Immirzi
parameter and the surface gravity corresponding to a null normal vector field.
We coment on the possible connection between processes involving vanishing of
punctures and Hawking radiation.  

\section{Counting of states}
We set our units such that $4\pi\gamma\ell_P^2=1$, where $\gamma$ is the 
so-called Barbero-Immirzi parameter involved in the quantization
and $\ell_P$ the Planck length. Equating the classical area $A$ of the
horizon to the eigenvalue (for a specific spin configuration of punctures on the
horizon) of the area operator we find
\beq A=2\sum_p\sqrt{j_p(j_p+1)}\;,\label{areacons}\eeq
where the $p$-th puncture carries a spin $j_p>0$, more accurately an irreducible
spin representation labelled by $j_p$, contributing a {\em quantum of area}
$2\sqrt{j_p(j_p+1)}$ to the total area eigenvalue.

Let the configuration be such that $s_j$ is the number of punctures carrying spin $j$. So
in (\ref{areacons}) the sum over punctures can now be replaced by the sum over spins
\beq A=2\sum_js_j\sqrt{j(j+1)}\;.\label{areacon2}\eeq
Such a spin configuration will be called {\em permissible} if it obeys
(\ref{areacon2}) together with the {\em spin projection constraint} which will be
introduced shortly.

\subsection{The combinatorial method}
First, we briefly review the calculation of \cite{gm} before going on to apply the
method to the counting of horizon states. Given a configuration labelled by $s_j$, 
different projections $m$ of $j$ give $\prod_j(2j+1)^{s_j}$ quantum states.
But the $s_j$s themselves can be chosen in $(\sum s_j)!/\prod s_j!$ ways
since the punctures are considered distinguishable. Therefore, the total number of
quantum states given by such a configuration $s_j$ is
\beq d_{s_j}={(\sum_js_j)!\over\prod_js_j!}\prod_j(2j+1)^{s_j}\;.\label{sconf}\eeq
To obtain the total number of states for all configurations (\ref{sconf}) is to be
summed over all configurations. We estimate the sum by maximizing $\ln d_{s_j}$ by
varying $s_j$ subject to (\ref{areacon2}). In the variation we assume that $s_j\gg
1$ for each $j$ and only such configurations dominate the counting. Such an
assumption breaks down if $A\sim o(1)$. The variational equation $\delta\ln
d_{s_j}=\lambda\delta A$, where $\lambda$ is the Lagrange multiplier, gives
\beq {s_j\over\sum s_j}=(2j+1)\,e^{-2\lambda\sqrt{j(j+1)}}\;.\label{snsol1}\eeq
Clearly, for consistency ({\em i.e.} summing both sides over all $j$), $\lambda$
must obey (cf. \cite{krasnov})
\beq 1=\sum_j(2j+1)\,e^{-2\lambda\sqrt{j(j+1)}}\;.\label{lambdae}\eeq

The counting however should also incorporate the spin {\em projection} constraint.
In order to implement this constraint the configuration must be given finer labels.
Let $s_{j,m}$ denote the number of punctures carrying spin $j$ with projection $m$.
With these new variables the area and the spin projection constraints take the
respective simple forms
\beq A=2\sum_{j,m}s_{j,m}\sqrt{j(j+1)}\;,\quad
0=\sum_{j,m}ms_{j,m}\;.\label{newc}\eeq
A configuration $s_{j,m}$ will be called permissible if it satisfies both of these
equations (\ref{newc}). The total number of quantum states for these configurations
is
\beq d_{s_{j,m}}={(\sum_{j,m} s_{j,m})!\over\prod_{j,m}s_{j,m}!}\;.\label{newco}\eeq
To obtain the dominant permissible configuration that contributes the largest number
of quantum states, we maximize $\ln d_{s_{j,m}}$ by varying $s_{j,m}$ subject to
(\ref{newc}). The result can be expressed in terms of two Lagrange multipliers
$\lambda,\alpha$:
\beq {s_{j,m}\over\sum s_{j,m}}=e^{-2\lambda\sqrt{j(j+1)}-\alpha
m}\;.\label{news}\eeq
Consistency requires that $\lambda$ and $\alpha$ be related to each other by $\sum_j
e^{-2\lambda\sqrt{j(j+1)}}\sum_me^{-\alpha m}=1$. In order that (\ref{news}) satisfies
the spin projection constraint we require
$\sum_je^{-2\lambda\sqrt{j(j+1)}}\sum_mme^{-\alpha m}=0$. This is possible if and
only if $\sum_mme^{-\alpha m}=0$ for each $j$, which essentially implies $\alpha=0$
(the value $2i\pi$ is excluded by positivity of $s_{j,m}$). Therefore,
the condition (\ref{lambdae}) on $\lambda$ remains unchanged. Note that each 
$s_{j,m}$ is proportional to the area $A$ because of the area constraint.

The total number of quantum states for all permissible configurations is clearly
$d(A)=\sum_{ s_{j,m}}d_{s_{j,m}}$. To estimate $d(A)$ we expand $\ln d$ around the
dominant configuration (\ref{news}), which we shall denote by $\bar s_{j,m}$. Thus
$\ln d=\ln d_{\bar s_{j,m}}-{1\over 2}\sum\delta s_{j,m}K_{j,m;j'm'}\delta
s_{j'm'}+o(\delta s_{j,m}^2)$ where $K$ is the symmetric matrix
$K_{j,m;j'm'}=\delta_{jj'}\delta_{mm'}/\bar s_{j,m}-1/\sum_{k,l}\bar s_{k,l}$. All
variations $\bar s_{j,m}+\delta s_{j,m}$ must satisfy the two conditions
(\ref{newc}) which yield two conditions $\sum\delta s_{j,m}\sqrt{j(j+1)}=0$ and
$\sum\delta s_{j,m}m=0$. Taking into account these equations we can express the
total number of states as
\beq d&=&d_{\bar s_{j,m}}\sum_{-\infty}^\infty e^{-{1\over 2}\sum \delta
s_{j,m}K_{j,m;j'm'}\delta s_{j'm'}}\cdot\nn
&&\delta(\sum\delta s_{j,m}\sqrt{j(j+1)})\;
\delta(\sum\delta s_{j,m}m)\nn
&=&Cd_{\bar s_{j,m}}\Big[\prod_{j,m}\sqrt A\,\Big]/A, \eeq
where $C$ is a constant independent of $A$. The denominator takes the particular
form because the two constraints remove two factors of $\sqrt A$, which would be
present otherwise in the Gaussian sum. It may be noted that $K$ has a zero
eigenvalue, but this is taken care of by the area constraint and all other
eigenvalues of $K$ are proportional to $1/A$. Inserting (\ref{news}) into
(\ref{newco}) and dropping factors of $o(1)$ we obtain
\beq d_{\bar s_{j,m}}=\frac{(\sum\bar
s_{j,m})^{1/2}}{\prod_{j,m}(2\pi\bar s_{j,m})^{1/2}}\,e^{\lambda A}\;.\eeq
Plugging these expressions into $d$ we finally obtain
\beq d={\alpha\over\sqrt A}\;e^{\lambda A}\;,\quad{\rm where}\;\alpha\sim o(1)\;,\eeq
leading to the formula \cite{gm}
\beq S=\lambda {A\over 4\pi\gamma\ell_P^2}-\frac12 \ln {A\over 4\pi\gamma\ell_P^2}
\label{s}\eeq
for entropy. The origin of the $\sqrt A$ in $d$ or $\half\ln A$ in $\ln d$ can be
easily traced in this approach: it is the condition $\sum ms_{j,m}=0$. This shows
that the coefficient of the log-correction is robust and does not depend on the
details of the configurations at all. It is directly linked with the boundary
conditions the horizon must satisfy.

\subsection{Application of the method to horizon states}
The above calculation was based on the understanding that $j$ is a relevant quantum
number. An alternative plan, adopted in \cite{ash, meissner}, is instead to count
the states of the horizon Hilbert space alone. For this purpose, one has to consider
the number $s_m$ of punctures carrying spin projection $m$, ignoring what spins $j$
they come from. One can distinguish between $s_j$ and $s_m$ from the context. It is
clear that
\beq s_m=\sum_js_{j,m},\quad j=|m|,|m|+1,|m|+2,... .\eeq
For the $s_m$ configuration the number of states is
$d_{s_m}=(\sum_{m}s_{m})!/\prod_{m}s_{m}!$ and the total number of states is
obtained by summing over all configurations. As in the earlier cases, the sum can be
approximated by maximizing $\ln d_{s_m}$ subject to the conditions (\ref{newc}). The
calculation resembles the previous one in spirit, but there are important
differences as discussed below. The constrained extremization conditions for
variation of $s_{j,m}$ are
\beq -\big[\ln{s_{m}\over\sum_ms_{m}}+2\lambda\sqrt{j(j+1)}+\alpha m\big]
=0.\label{consm}\eeq
Clearly, the above equations cannot hold for arbitrary $j$ even for a fixed $m$,
because inconsistencies will arise for nonzero $\lambda$. In fact, for any fixed $m$
the above equation admits at most one $j$ -- let us denote it by $j(m)$.
For $j\neq j(m)$, the first derivative becomes nonzero. Such a situation can arise
if and only if $\ln d_{s_m}$ is maximized at the boundary (in the space of all
permissible configurations) for all $j\neq j(m)$ and at an interior point for
$j=j(m)$. This means that for the dominant configuration $s_{j,m}=0$ for all $j\neq
j(m)$: the corresponding first derivative is then only required to be zero or
negative because in any variation $s_{j,m}$ can only increase from its zero value.
Thus, $s_m=s_{j(m),m}$ for the dominant configuration.

Then (\ref{consm}) gives
\beq {s_{m}\over\sum_{m}s_{m}}=e^{-2\lambda\sqrt{j(m)(j(m)+1)}-\alpha
m}\label{consf}\;.\eeq
As before, $\alpha=0$ in order that the dominant configuration satisfies the spin
projection constraint. The parameter $\lambda$ is determined by a consistency
condition involving $j(m)$. Since the entropy increases with $\lambda$, and lower
$j(m)$ gives higher $\lambda$, the maximum entropy is obtained when $j(m)$ is
minimum, {\em i.e.,} $j(m)=j_{\rm min}(m)$, the minimum value for $j$ for a given
$m$. For all $m\neq 0$, we have $j_{\rm min}(m)=|m|$. But for $m=0$, we must have
$j_{min}(m)=1$, since $j=0$ is excluded.

The configuration (\ref{consf}) with $j(m)=j_{\rm min}(m)$ implies that
the entropy is given by (\ref{s}) in terms of $\lambda$, which is now determined
by the altered consistency relation
\beq 1=\sum_{j\neq 1}2e^{-2\lambda\sqrt{j(j+1)}}+3e^{-2\lambda\sqrt{2}},\eeq
where each $j\neq 1$ is associated with $m=\pm j$ only, but $j=1$ also has $m=0$.
Note that for $\lambda$ zero or negative, such relations would be impossible to
satisfy, hence no such solutions exist.

This equation for $\lambda$ differs from that of \cite{meissner} in
allowing $m=0$ for $j=1$ and thus yields a slightly greater value $0.790$
instead of $0.746$. The
difference arises because we have used the area constraint directly,
using the definition of the area involving $j$.
In contrast, \cite{meissner} used an inequality involving $m$,
\beq A\geqslant 2\sum_ms_m\sqrt{|m|(|m|+1)},\eeq
which can be derived on the basis
of the inequality $j\geqslant|m|$, but is not saturated for punctures with $j=1,m=0$,
which the maximization conditions allow.
The value $0.790$ of $\lambda$ is naturally less than the value $0.861$ obtained by
taking both $j$ and $m$ to be relevant quantum numbers \cite{gm}.

It is to be noted that our counting of horizon states allows {\em three}
spin states for $j=1$ and is thus consistent with the general ideas in
\cite{dreyer} which reported an intriguing connection between the spin
degeneracy and an observed factor of $\ln 3$ occurring in the classical
quasinormal modes of black holes. \cite{dreyer} recommends {\em only}
$j=1$, which could be accommodated by setting $s_{j,m}=0$ for all
$m$ except 0 and 1.
Our earlier calculation of bulk states
\cite{gm} was also consistent with \cite{dreyer} and coincides
with the present calculation for $j=1$. In contrast, the
counting of \cite{meissner} allows only {\em two} projection states
for  $j=1$ and is therefore, inconsistent with \cite{dreyer}.

\section{Towards the definition of a temperature}

First we make some comments on the interpretation of the laws of the mechanics 
of a weakly isolated horizon (WIH) as thermodynamic laws. 

{\tt (1)} The zeroeth law of WIH states that the surface gravity 
$\kappa_{(\ell\,)}$ associated with each `fixed' null normal vector 
$\ell^a$ (which generates the WIH) is constant on the horizon. 
However for a given isolated horizon, $\ell^a$ is fixed only up to a 
constant rescaling. Under such a rescaling $\ell^a\mapsto c\ell^a$, 
where $c$ is a positive number, both $\kappa_{(\ell\,)}$ and the 
`horizon-mass' $M_{(\ell\,)}$ (which also depends on the choice of 
$\ell^a$, see \cite{ashiso} for details) are rescaled by the same constant 
$c$, whereas the horizon-area $A$ does not alter. In fact the first law of 
a non-rotating WIH, which states that the change of the horizon-mass
\beq \Delta M_{(\ell\,)}={\kappa_{(\ell\,)}\over 8\pi G}\;\Delta A\;,\label{1stl}\eeq
where $\Delta A$ is the associated change of the horizon-area, depends 
explicitly on $\ell^a$ (although the above scaling argument show that the 
form (\ref{1stl}) is independent of $\ell^a$). Thus, both zeroeth and 
first laws of WIH make an explicit reference to a `fixed' null normal 
vector field $\ell^a$. This fact is to be kept in mind whenever we draw 
analogies between a WIH and a thermodynamic system. Unless some $\ell^a$ 
is fixed, confusions will arise in the thermodynamic interpretation of a WIH. 

{\tt (2)} This is regarding the quantum statistical mechanics of a WIH. Now 
that there is a quantum mechanical entropy of a WIH, we have definite quantum 
states of a WIH. However, a realistic statistical interpretation of a WIH, even
as a microcanonical ensemble, requires states of the `environment', {\it viz.,}
the states of the bulk of the spacetime of which the WIH is a subsystem, the
bulk and the WIH together forming an isolated system.
The microcanonical ensemble assumes a weak interaction between the bulk and 
the horizon such that the horizon-area $A$ is constant (more precisely, it 
lies in a small interval $[A-\epsilon,A+\epsilon]$ where $\epsilon\ll A$). 
The trace over the bulk spacetime states provides a density matrix for the WIH
(which for a microcanonical ensemble is trivial, proportional to the identity
matrix). This is related to the comment made in the introduction that a quantum
WIH can never be fully described by surface states alone, the bulk states act
like a heat-bath as indicated above.

It is not at all difficult to arrive at an expression of a microcanonical
temperature based on the formal analogy with thermodynamics. We already
found that for a fixed $\ell^a$ the surface gravity $\kappa_{(\ell\,)}$
should be related to the temperature and the first law (\ref{1stl}) is to
be interpreted as the first law of thermodynamics. Since the entropy is
given by (\ref{s}), its variation is (ignoring the log-correction for now)
$\Delta S=\lambda\Delta A/4\pi\gamma\ell_P^2$ and equating $T_{(\ell\,)}
\Delta S$ with the RHS of (\ref{1stl}) we get an expression
\beq T_{(\ell\,)}=\frac{\hbar\gamma\kappa_{(\ell\,)}}{2\lambda}\;.\label{temp}
\eeq
This is the microcanonical temperature of a WIH having a fixed null normal
vector field.

To interpret the microcanonical ensemble as a canonical or a grand-canonical
ensemble we need to allow interactions between the WIH and the bulk. 
For each permissible configuration $s_j\equiv\sum_ms_{j,m}$ the area spectrum is
$A=8\pi\gamma\ell_P^2\sum s_j\sqrt{j(j+1)}$. Now imagine a quantum mechanical 
process that changes the configuration $s_j$ to another permissible 
configuration $s_j+\Delta s_j$, that causes the area to change by
$\Delta A=8\pi\gamma\ell_P^2\sum\Delta s_j\sqrt{j(j+1)}$. 
(This change $\Delta s_j$ should not be confused with $\delta s_j$ we used 
earlier. Here permissibility of the new configuration $s_j+\Delta s_j$ does 
not imply $\sum\Delta s_j\sqrt{j(j+1)}=0$: while the permissibility of $s_j$ 
is associated with the area $A$, the one of $s_j+\Delta s_j$ is associated 
with the area $A+\Delta A$, where $\Delta A$ is a physical change of area.)
Thus, from (\ref{1stl}) we obtain
\beq \Delta M_{(\ell\,)}=\hbar\kappa_{(\ell\,)}\gamma\sum\Delta 
s_j\sqrt{j(j+1)}\;.\label{delm}\eeq
This is a key result showing how the mass/energy of the WIH can leak to the
bulk of the spacetime. This involves the creation and annihilation of punctures.
In a microcanonical ensemble these processes take place only under the strict 
permissibility conditions (which basically ensure that the area and
the energy cannot change). But in a canonical or grand-canonical ensemble 
these restrictions are to be removed. A detailed study is required in this 
direction to interpret the temperature (\ref{temp}) in a canonical ensemble.

Since a WIH involves an infinite family of null normal vectors, it also admits 
an infinite family of corresponding temperatures, fixed for each fixed $\ell^a$. 
Moreover, (\ref{temp}) shows that the relation $\gamma\pi=\lambda$ which yields 
the semiclassical expression of entropy, also gives the semiclassical 
expression of temperature $T_{(\ell\,)}=\hbar\kappa_{(\ell\,)}/2\pi$. 
It is interesting to ask
what alteration the log correction to the entropy (\ref{s})
induces in the temperature. A simple calculation shows that $T_{(\ell\,)}=
(\hbar\kappa_{(\ell\,)}\gamma/2\lambda)(1+2\pi\gamma\ell_P^2/\lambda A)$. 
So while the entropy receives a universal log-correction, the temperature 
is corrected only by a power-law. Unlike the case of the entropy, the 
coefficient of the power-law correction is not universal - it depends 
on the underlying quantum theory. However, the value of $\gamma$ that 
gives the semiclassical sector of
quantum gravity also makes the coefficient independent of $\lambda$.

\section{Discussion}
We have followed the combinatorial approach of \cite{gm} to count
horizon states and have found that there are more of these than
indicated by the approximate analysis of \cite{meissner}. The increased
number is of course still not as large as the total number of microscopic
states found in \cite{gm} where not only $m$ but also $j$ was regarded
as a relevant label for a microscopic state. However, the correction
brings the number of states distinguished by $m$ closer to the number
of states labelled by $j,m$ and also makes it consistent with
\cite{dreyer}. 

Thereafter we have sought to introduce a temperature
corresponding to each choice of the null normal vector field $\ell^a$.
The discussion in the previous section suggests that 
the area ensemble may be regarded as an energy
ensemble for each fixed $\ell^a$. Standard statistical mechanical arguments
then may permit us to view the microcanonical ensemble as a canonical or
grand-canonical ensemble. At thermal equilibrium the quantum mechanical process
changing the horizon-area suggests the following picture: quantum states 
associated with the punctures get annihilated from the surface Hilbert space 
by transforming into bulk states. If the bulk is taken to be asymptotically 
flat then such bulk states appear to be the usual Fock states. Reversibly, the 
Fock states from the bulk of the spacetime must
transform into the puncture-states and these two processes must take place at 
the same rate. These processes are quite analogous, though not identical, to 
the particle creation and annihilation processes we encounter in 
quantum field theories. In quantum field theories in flat space the creation 
and destruction of one-particle states are performed by certain linear operators 
in the Fock space. Furthermore, such one-particle
states are labelled by their energy and momenta, so a fixed stationary background 
metric is required. However here we are considering creation and annihilation
of punctures in changing $s_j$ to $s_j+\Delta s_j$. Moreover, no
background metric is present. Punctures are also labelled by the spin quantum 
numbers. The linear operators that can create or destroy punctures
should be related to the spin-raising and spin-lowering operators in the bulk Hilbert
space of loop quantum gravity. Such operators have indeed been constructed while 
obtaining the area-spectrum \cite{al}. For the time being, it is an open problem 
to show that such processes exist in the Hilbert space within the framework of 
loop quantum gravity. Of course, the bigger question is whether, when the 
reverse process (bulk states to surface states) 
is ignored, the forward process (surface states to bulk states) appears as 
black-body radiation.

One can also arrive at a generalized
statement of the second law that the combined entropy of the horizon and the bulk
does not decrease. While the microscopic degrees of freedom associated with the
horizon are punctures, those of the bulk remain the standard matter and field
particles. In order that a thermal equilibrium is reached, these two degrees of
freedom must transform into each other continuously. It remains to be seen how
such a picture emerges in quantum geometry.


\end{document}